\documentclass[12pt,preprint]{emulateapj}      
\usepackage{epsfig}
\slugcomment{ApJ Accepted}

\newcommand{\gray }{$\gamma$-ray\ }

\newcommand{\piodecay}{$\pi^0$-decay\ }
\newcommand{\hi}{H {\sc i}}

\newcommand{\EB}{EGRB}

\newcommand{\intensityunits}{$\times10^{-6}$~cm$^{-2}$~sr$^{-1}$~s$^{-1}$\ }

\newcommand{\regionG}{$360^\circ<l<0^\circ,10^\circ<|b|<80^\circ$}
 
\newcommand{\fw}{53mm}
\newcommand{\fwa}{71mm}
\newcommand{\fwb}{160mm}

\newcommand{\app}{Astropart.\ Phys.}
\newcommand{\na}{New A}

\newcommand{\pubjournal}[5]{#4, #1, #2, #3}
\newcommand{\pubjournala}[5]{#4, #1, #3}

\shorttitle{Extragalactic Diffuse \gray Background}
\shortauthors{Strong, Moskalenko, \& Reimer}
\begin{document}
\title{A new determination of the extragalactic diffuse gamma-ray background\\ from EGRET data}

\author{Andrew W.~Strong}
\affil{Max-Planck-Institut f\"ur extraterrestrische Physik,
   Postfach 1603, D-85740 Garching, Germany; aws@mpe.mpg.de}

\author{Igor V.~Moskalenko\altaffilmark{1}}
\affil{NASA/Goddard Space Flight Center, Code 661, Greenbelt, MD 20771; 
moskalenko@gsfc.nasa.gov}
\altaffiltext{1}{Joint Center for Astrophysics, University of Maryland, 
   Baltimore County, Baltimore, MD 21250}

\and

\author{Olaf Reimer}
\affil{Ruhr-Universit\"at  Bochum, D-44780 Bochum, Germany; olr@tp4.ruhr-uni-bochum.de}


\begin{abstract}
We use the GALPROP model for cosmic-ray propagation to obtain a new
estimate of the Galactic component of gamma rays, and show that away
from the Galactic plane it gives an accurate prediction of  the
observed EGRET intensities in the energy range 30 MeV -- 50 GeV.  On
this basis we re-evaluate the extragalactic gamma-ray background.  We
find that for some energies previous work underestimated the Galactic
contribution at high latitudes and hence overestimated the
background. Our new background spectrum shows a positive curvature
similar to that expected for models of the extragalactic emission
based on the blazar population.

\end{abstract}



\section{Introduction}
The extragalactic diffuse \gray background emission (\EB) is a
superposition of all unresolved sources of high energy \gray emission
in the Universe.  Active galactic nuclei (AGN) are the dominant class
of \gray emitters known to emit up to the highest energies. There is a
consensus that a population of unresolved AGN may contribute to the
\EB, however predictions range from 25\% up to 100\%.  Contributions
from other extragalactic sources have been plausibly suggested: galaxy
clusters \citep{ensslin97}, energetic particles in the shock waves
associated with  large scale cosmological structure formation
\citep{loeb00,miniati02}, or distant \gray burst events.  Potentially,
if reliably derived, the \EB\ can also provide important information
about the phase of baryon-antibaryon annihilation \citep{gao,dolgov},
evaporation of primordial black holes \citep*{hawking,maki},
annihilation of so-called weakly interacting massive particles (WIMPs)
\citep*{jkg}, and extragalactic IR and optical photon spectra
\citep{stecker}.

The \EB\ is the component of the diffuse emission which is most
difficult to determine.  Its spectrum depends much on the adopted
model of the Galactic background which itself is not yet firmly
established.  The isotropic, presumably extragalactic component of the
diffuse \gray\ flux was first discovered by the SAS-2 satellite and
confirmed by EGRET \citep{thompson82,sreekumar98}. However, it is not
correct to assume that the isotropic component is wholly extragalactic
since even at the Galactic poles it is comparable to the Galactic
contribution from inverse Compton scattering of the Galactic plane
photons and CMB \citep*{SMR00,MS00}.  The determination of the \EB\
is thus model-dependent and influenced by the adopted size of the
Galactic halo, the electron spectrum there, and the spectrum of
low-energy background photons which must be derived independently.

Extensive work has been done \citep{sreekumar98} to derive the
spectrum of the \EB\ based on EGRET data.  The relation of modelled
Galactic diffuse emission to total measured diffuse emission was used
to determine the \EB, as the extrapolation to zero Galactic
contribution.  The derived spectral index $-2.10\pm0.03$ appears to be
close to that of \gray blazars.

\citet{dixon98}, using an model-independent approach, also
concluded that the \EB\ will be affected
by  a significant contribution from a Galactic halo component.

In a companion paper \citep*{SMR04} we use the GALPROP code to infer a
new model for  Galactic diffuse continuum $\gamma$-rays.  This model
reproduces successfully diffuse \gray emission from the entire sky and
gives a good linear prediction for observed vs.\ predicted \gray
intensities.  In view of the success of this model, we use it as the
basis of a new determination of the \EB\ using EGRET data from 30 MeV
to 50 GeV.  The GALPROP propagation code and previous results are
described elsewhere \citep[and references therein]{M02,SMR04}.

\section{The procedure}

\subsection{EGRET data}
The details of the procedure of handling the EGRET data and
convolution procedure are described in detail elsewhere \citep{SMR04};
here we provide a brief summary.

We use the EGRET counts and exposure  all-sky maps in Galactic
coordinates with 0.5$^\circ$ binsize, as in \citet{SMR00}.  The
sources of the 3EG catalogue have been removed  by the procedure
described in \citet{SMR00}, fully consistent with the 3EG  point
source listings.  Apart from the most intense sources, the removal of
sources has little influence on the comparison with models if
sufficiently large sky segments are investigated.  For the spectra,
the statistical errors on the EGRET data points are very small since
the regions chosen have large solid angle; the systematic error
dominates and we have adopted values in the range 10--30\% depending
on energy as shown in Table~\ref{estimates_of_EB}, cf.\ 15\% adopted by
\citet{sreekumar98} and \citet{esposito}.  The predicted model skymaps are
convolved with the EGRET point-spread  function as described in
\citet{SMR00}.  Here we use additional EGRET data in the energy
ranges 10--20, 20--50 and 50--120 GeV. Because the instrumental
response of EGRET  determined at energies above 10 GeV is less certain
compared to energies below 10 GeV, it is required to account for
additional systematic uncertainties. In particular the EGRET effective
area can only be deduced by  extrapolation from the calibrated
effective area at lower energies \citep{Thompson1993a}.  We
accordingly adopt values of 0.9, 0.8, and 0.7 times the 4--10 GeV
effective area,  respectively.

\subsection{Optimized model for the Galactic diffuse emission}
In the companion paper \citep{SMR04} we compared a range of models of
\emph{Galactic}  diffuse emission, based on our CR propagation code
GALPROP, with data from the Compton Gamma Ray Observatory. There we
exploit the fact that the models predict quite specific behaviour for
different sky regions and this provides a critical test: the
``correct'' model should be consistent with the data in \emph{all}
directions.   We show that a new model, with moderate changes of
electron and nucleon spectra relative to the ``conventional'' model,
can well reproduce the \gray data and is compatible with locally
observed particle spectra considering the expected level of spatial
fluctuations in the Galaxy.  The \gray data comparisons were extended
over the entire sky and to 100 GeV in energy.  We also exploited the
recent improved measurements of the local proton, Helium, as well as
antiproton, and positron spectra which are used as constraints on the
proton spectrum in distant regions.

To fit the Galactic diffuse emission we used
7 test regions covering the sky \citep[for
details see][]{SMR04}. The model uses the proton (and He) spectral
shape at high energies derived from the local data. The 
secondary antiproton and
positron data tracing the \emph{proton} spectrum on a large scale provide an
important constraint on the \emph{intensity} normalization of the
\emph{average} spectrum. The
adopted \emph{average electron} spectrum resembles the local one
renormalized upwards by a factor of $\sim$4 and consistent with
synchrotron index measurements.  The adopted electron and nucleon
spectra are compatible with the direct measurement considering
fluctuations due to energy losses and the stochastic character of cosmic
ray sources and propagation.

The optimized model fits the observed Galactic diffuse \gray spectra
in all test regions in the energy range 30 MeV -- 100 GeV.  The
proposed scenario implies a substantial contribution from IC at all
energies, but especially below 100 MeV and above 1 GeV. Also IC
dominates at latitudes $|b|>10^\circ$ at all energies. The agreement
of longitude profiles with the EGRET data is generally good
considering that the model does not attempt to include details of
Galactic structure.  The agreement of latitude profiles with EGRET is
also good, in particular the reproduction of the high-latitude
variation confirms the importance of the IC component which is much
broader than the gas-related \piodecay and bremsstrahlung
emission. The outer Galaxy latitude profiles are in excellent
agreement with the data.

\begin{figure*}[!t]
\centering 
\includegraphics[width=\fw]{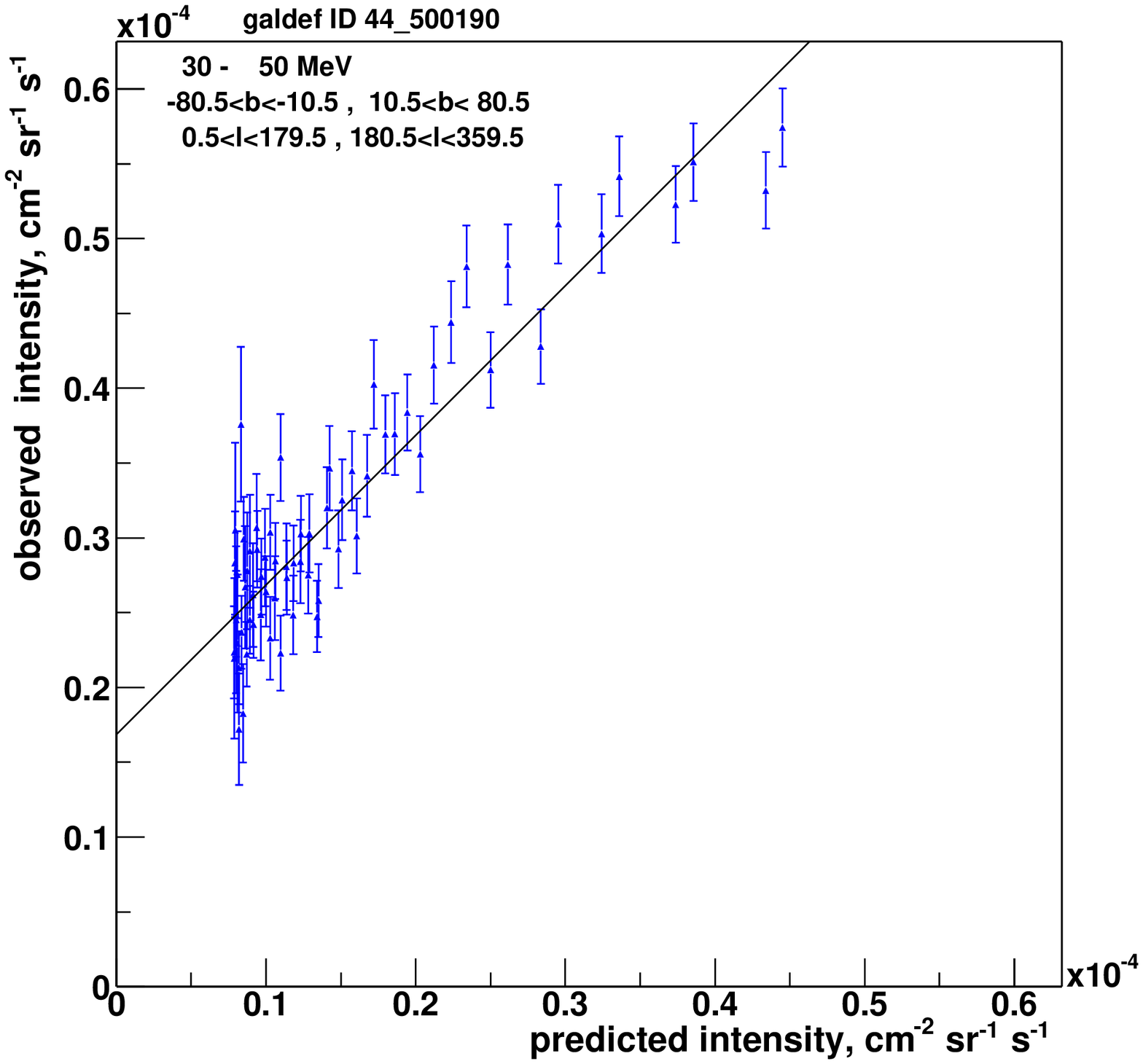}
\includegraphics[width=\fw]{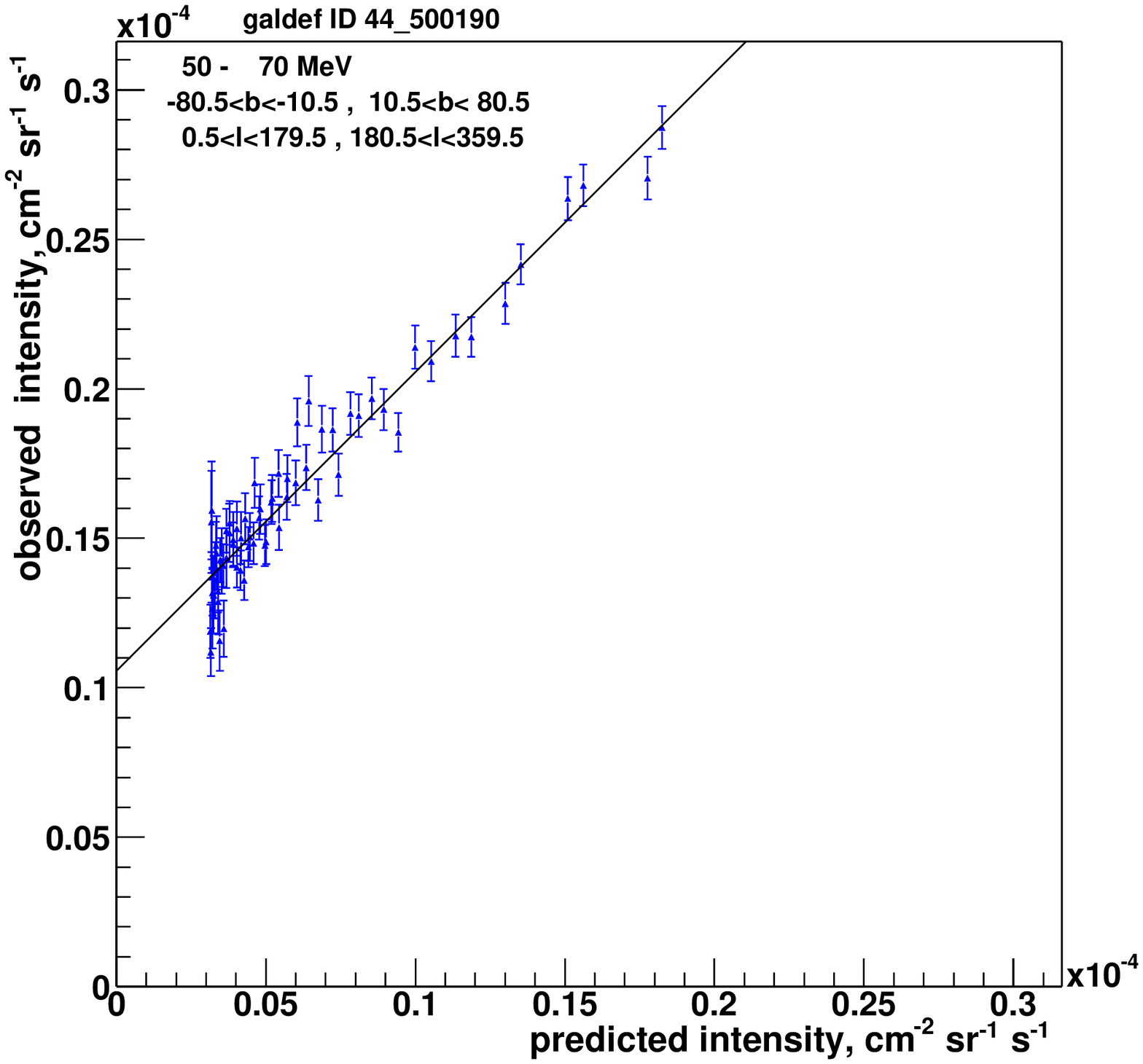}
\includegraphics[width=\fw]{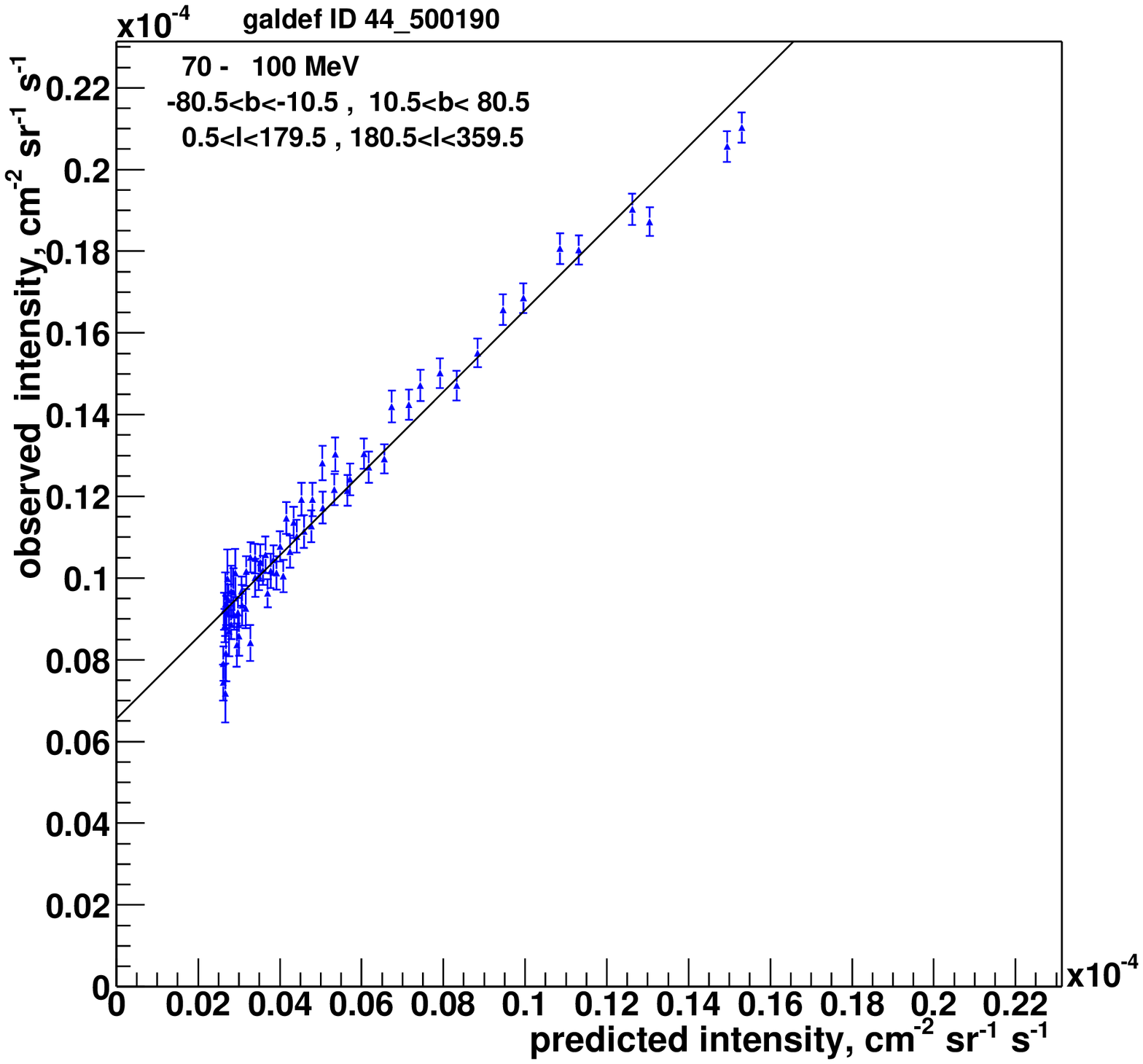}
\includegraphics[width=\fw]{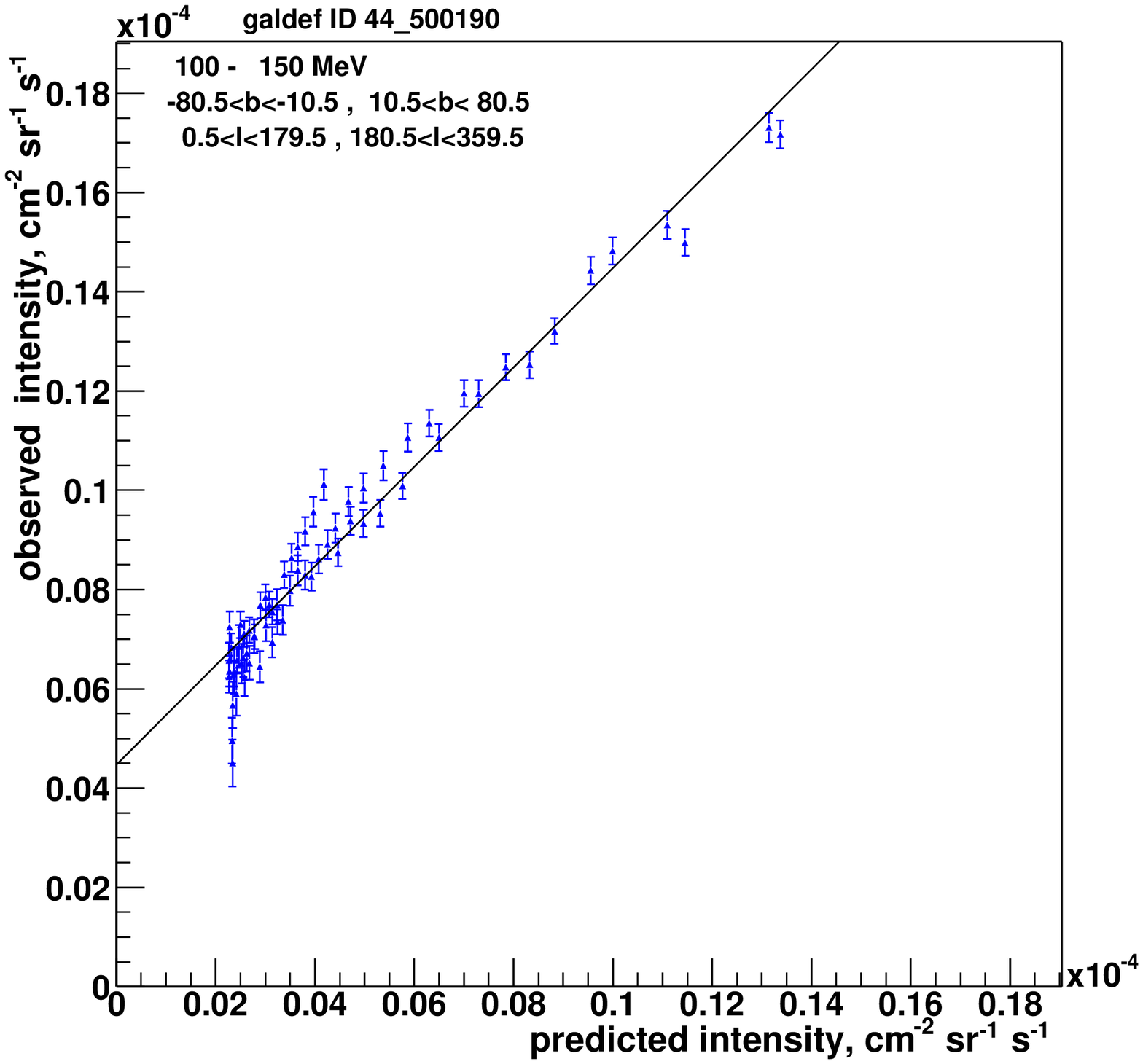}
\includegraphics[width=\fw]{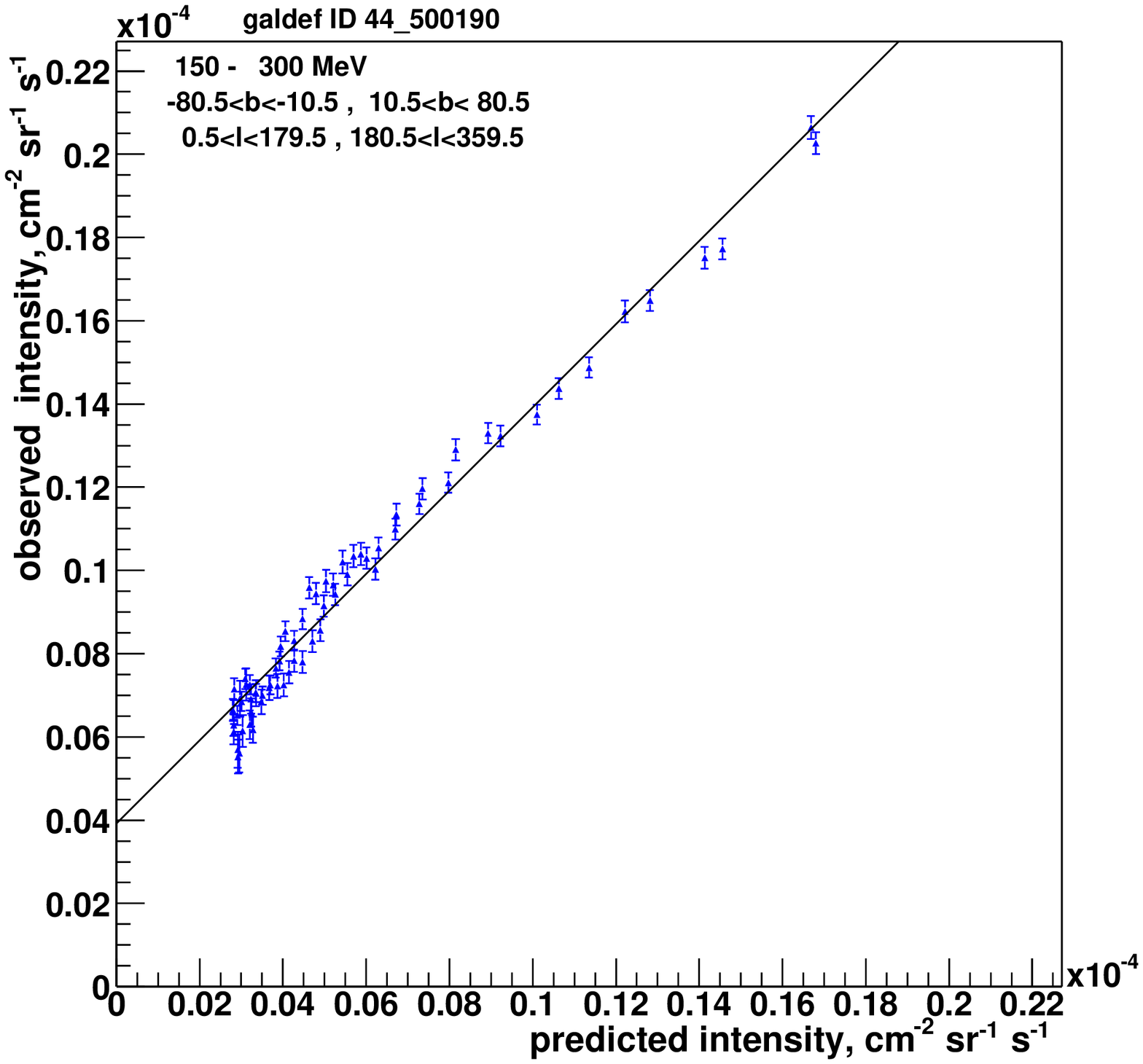}
\includegraphics[width=\fw]{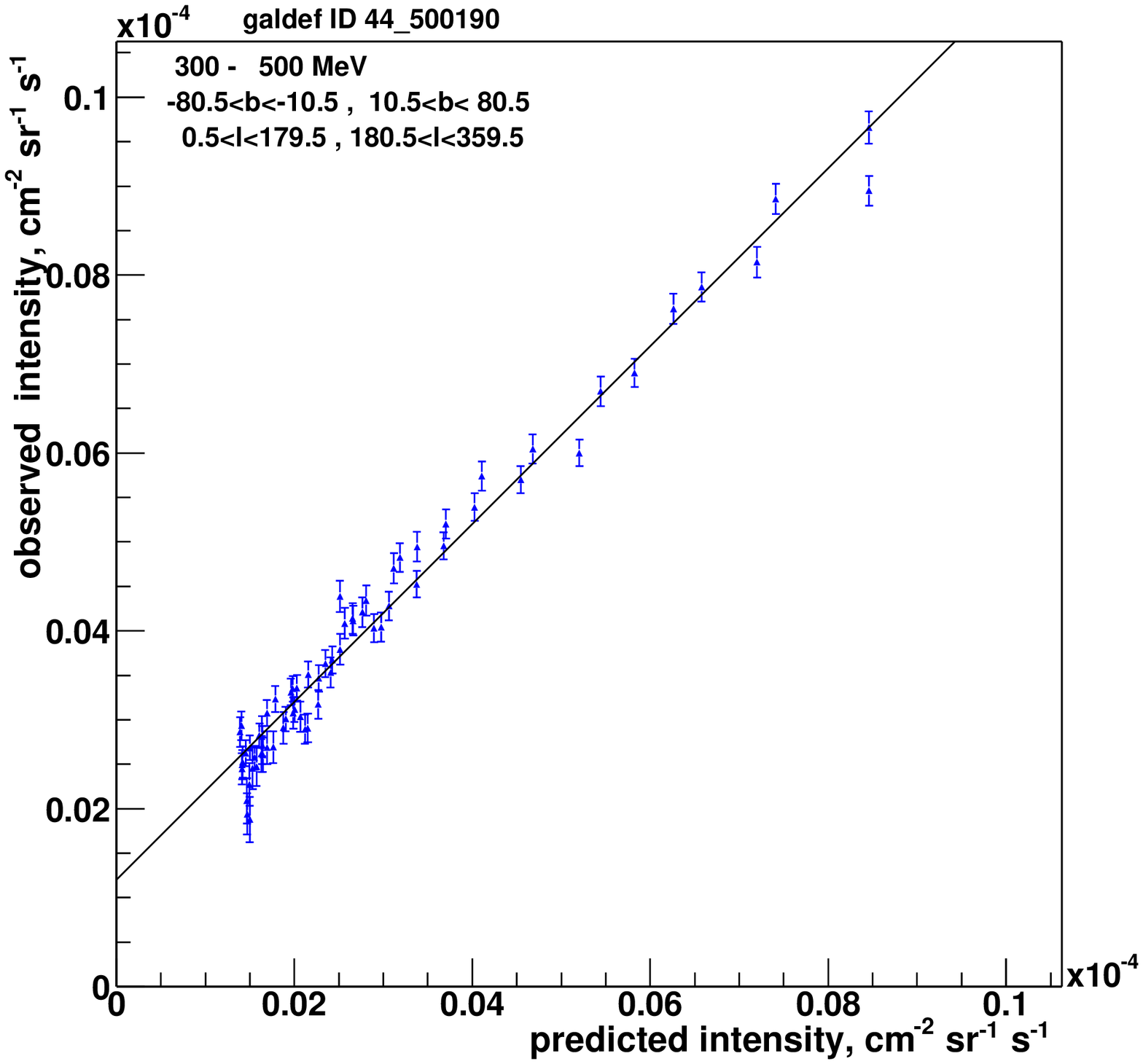}
\includegraphics[width=\fw]{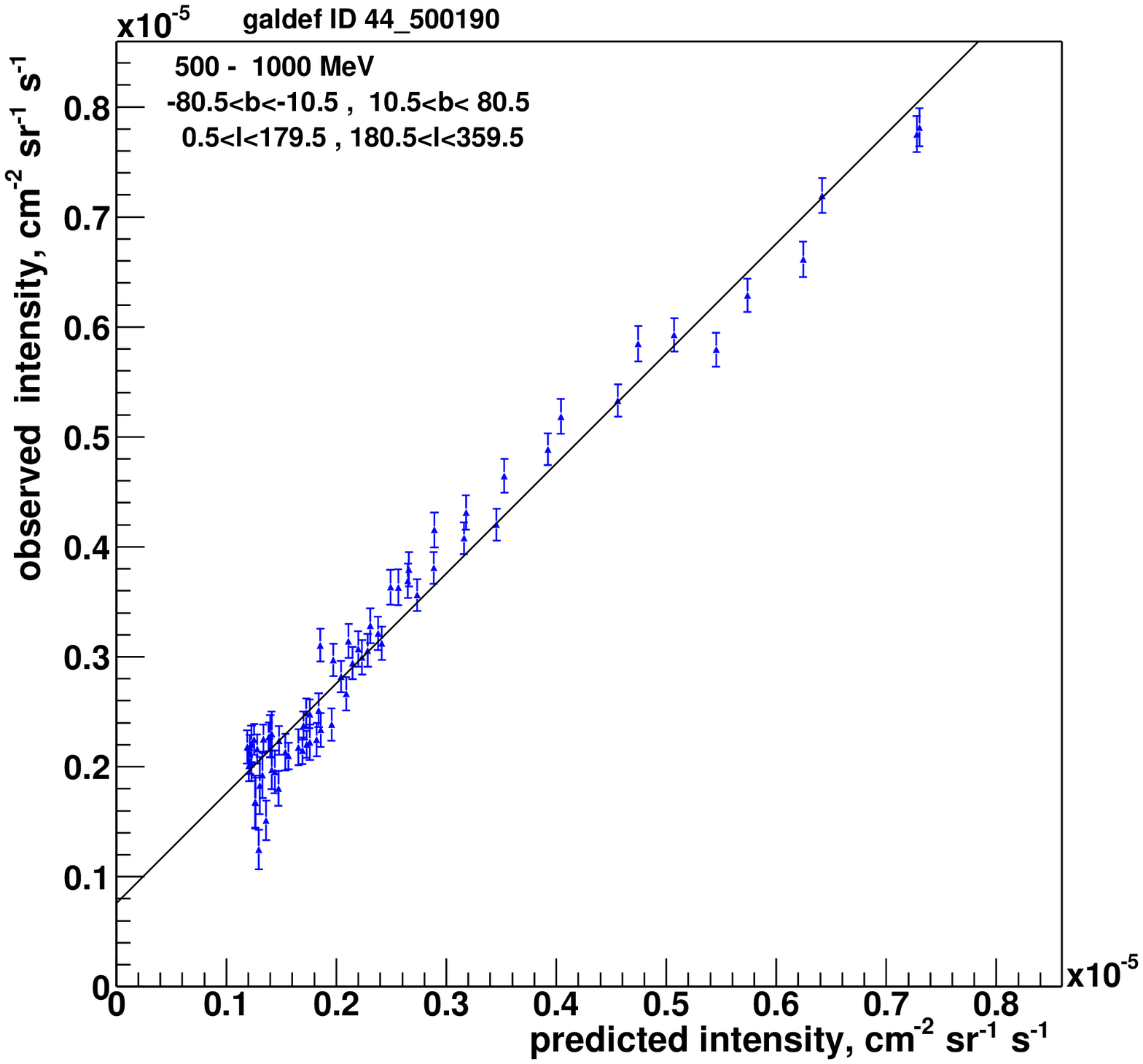}
\includegraphics[width=\fw]{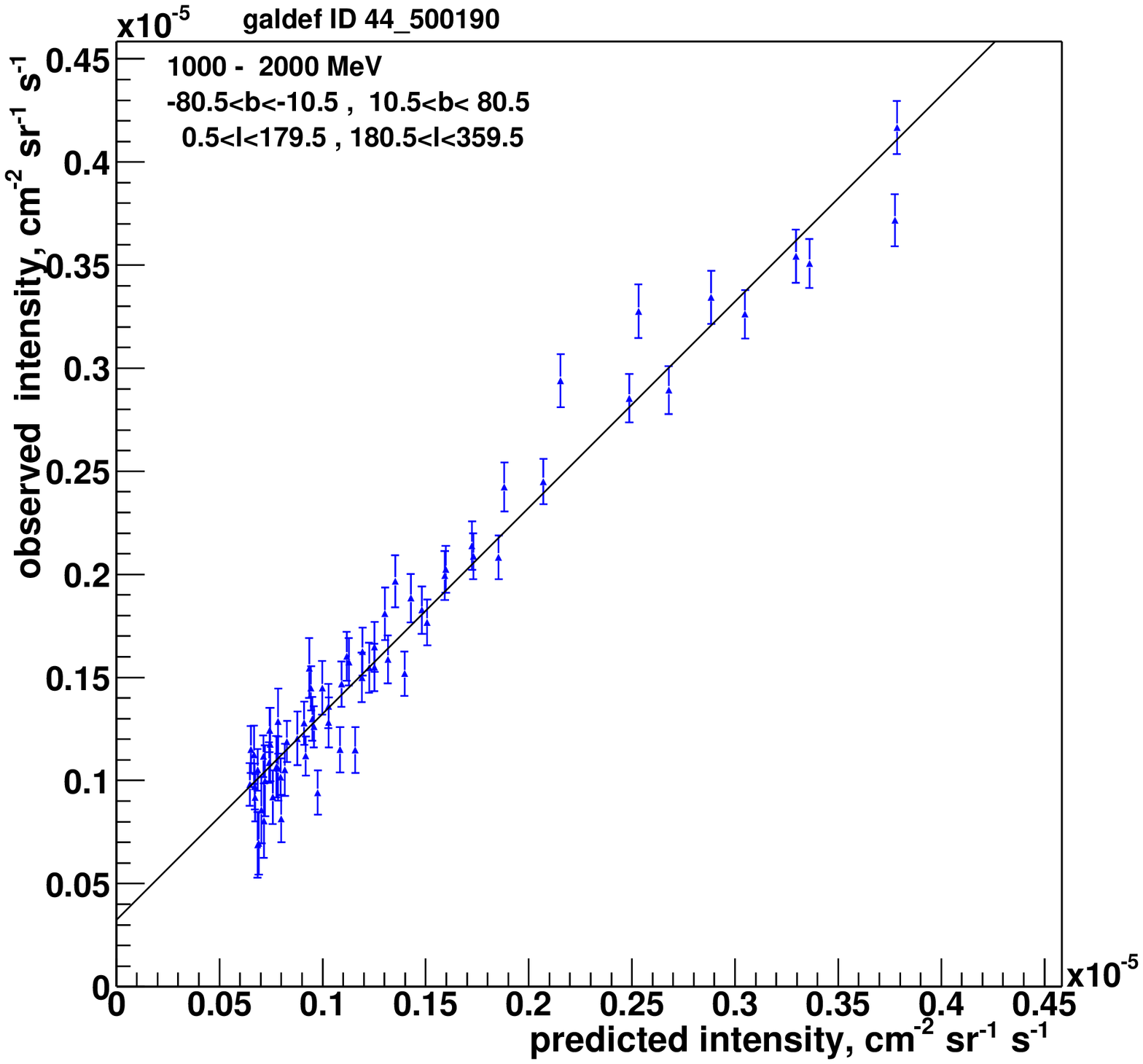}
\includegraphics[width=\fw]{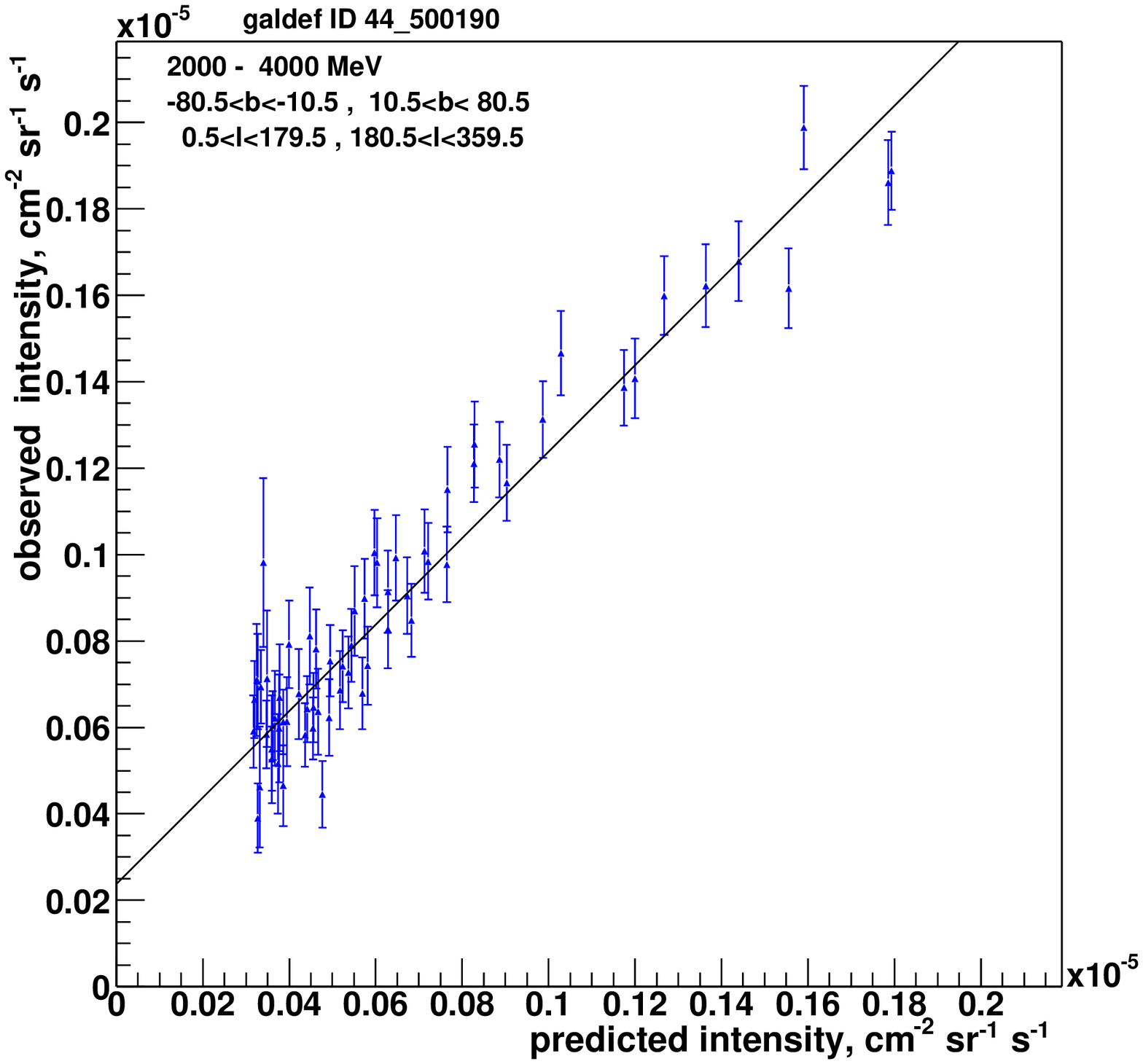}
\includegraphics[width=\fw]{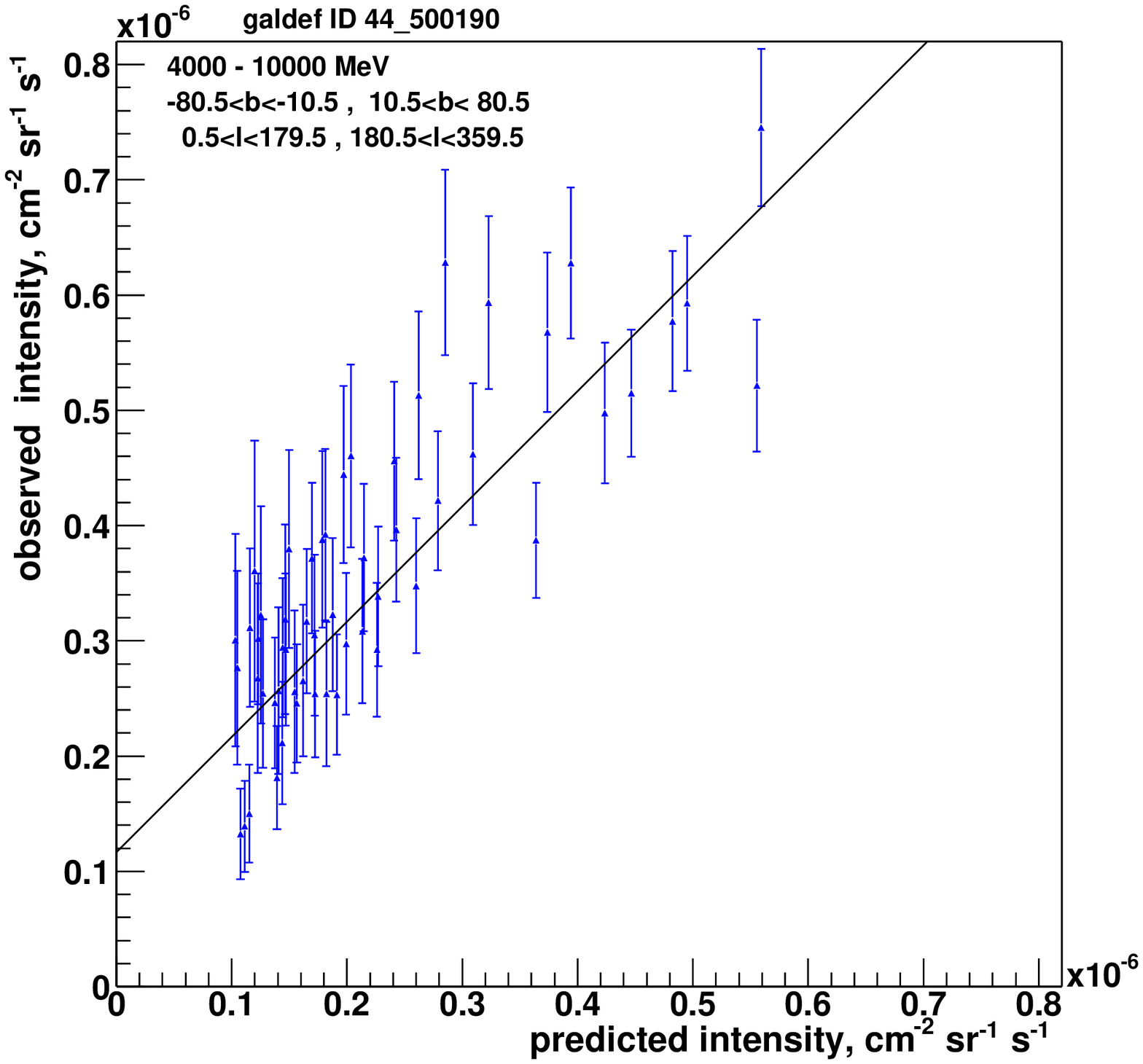}
\includegraphics[width=\fw]{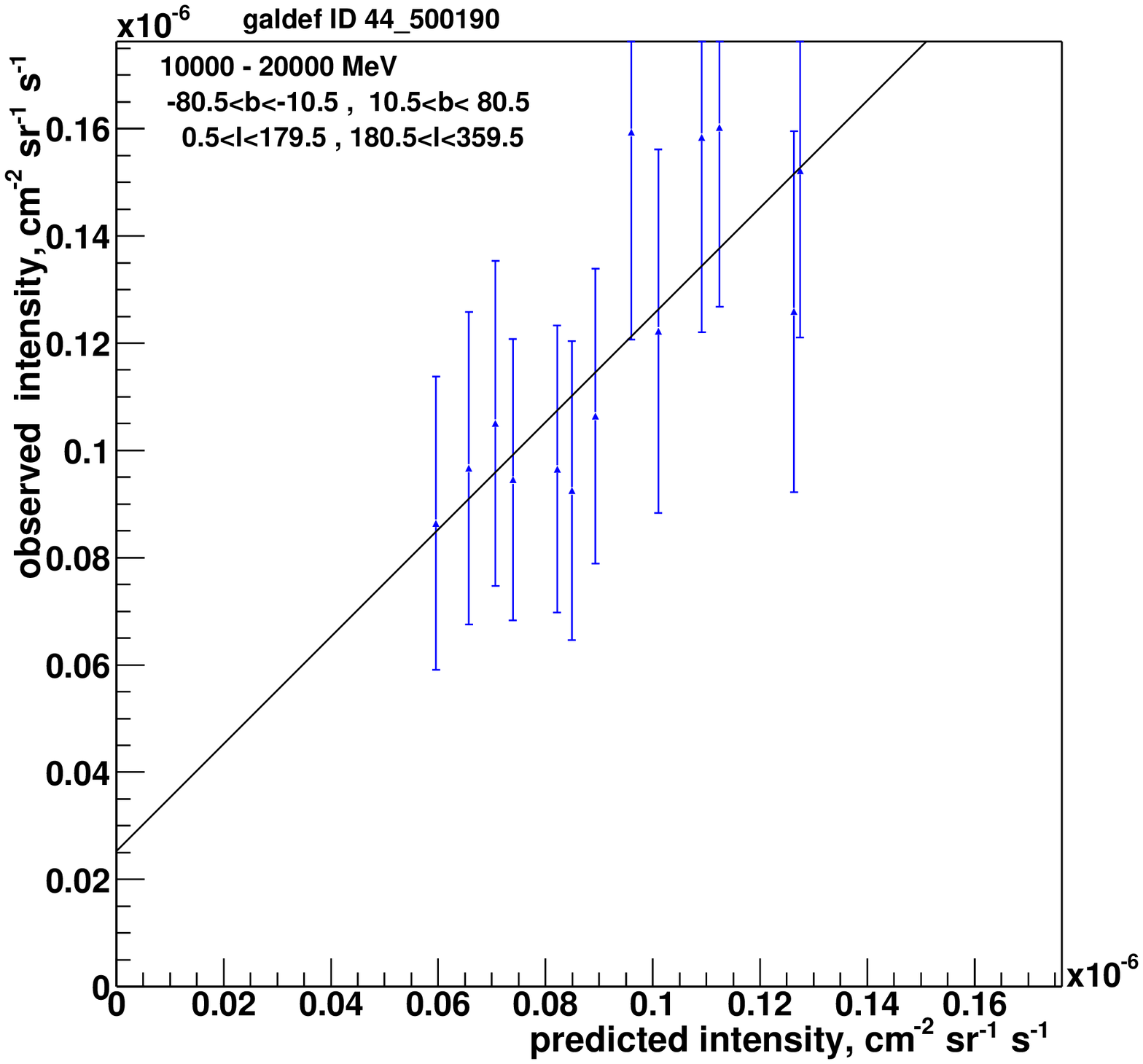}

\caption{Observed versus predicted intensities for region \regionG\
(GALPROP model ID 500190). Predicted is Galactic only, intercept is \EB.
}
\label{gamma_pred_obs_500190}
\end{figure*}

\placefigure{gamma_pred_obs_500190}

\subsection{Method}
Since despite its success  the model is nevertheless not exact the
best approach is to fit the observed intensities with a free scaling
factor, so that the \EB\ is determined as the intercept, thus removing
any residual uncertainty in the absolute level of the Galactic
components. This is the same method as in \citet{sreekumar98}, the
difference lies in the model.  To reduce the effects of Galactic
structure, point sources etc. the fits are made excluding the plane,
using the sky region  \regionG; ideally both IC and gas-related
components would be left free but they are difficult to separate
statistically at high latitudes, so we make a linear fit to the total
IC+$\pi^0$-decay+bremsstrahlung, with the scaling factor and \EB\ as
parameters. The fit and errors are based on a simple $\chi^2$
analysis, with ($l,b$) bins $360^\circ\times2^\circ$ to obtain
sufficient statistics (at least 10 counts per bin were required).  For
comparison we also made fits to the entire sky; in this case IC and
gas-related contributions are easily separated, so that fits with both
components free can be made in addition to fits to the total.  The
fits in the two  regions then give some indication of the
model-dependent systematic error in our \EB\ estimates.

\placetable{scaling_factors_EB}

\begin{deluxetable*}{ccccc}[!t]
\tablecolumns{5}
\tablewidth{0mm}
\tabletypesize{\footnotesize}
\tablecaption{ Scaling factors of model components
corresponding to fits in Table~\ref{estimates_of_EB}. 
\label{scaling_factors_EB}}
\tablehead{
\colhead{} &
\colhead{$10^\circ<|b|<80^\circ$,} &
\multicolumn{3}{c}{All-sky}\\ \cline{3-5}
\colhead{Energy, MeV} &
\colhead{total}&
\colhead{Total}&
\colhead{IC} &
\colhead{Gas} 
}
\startdata
30--50      &$ 1.008\pm 0.04  $&$1.07\pm0.017$ &$ 0.93  \pm 0.09 $&$ 1.31 \pm 0.16 $\\
50--70      &$ 0.864\pm 0.02  $&$0.88\pm0.009$ &$ 0.66  \pm 0.05 $&$ 1.13 \pm 0.06 $\\
70--100     &$ 0.872\pm 0.01  $&$0.84\pm0.005$ &$ 0.82  \pm 0.03 $&$ 0.86 \pm 0.02 $\\
100--150    &$ 0.851\pm 0.01  $&$0.82\pm0.004$ &$ 0.85  \pm 0.03 $&$ 0.80 \pm 0.01 $\\
150--300    &$ 0.874\pm 0.01  $&$0.83\pm0.003$ &$ 0.95  \pm 0.02 $&$ 0.79 \pm 0.01 $\\
300--500    &$ 0.965\pm 0.01  $&$0.90\pm0.004$ &$ 1.14  \pm 0.04 $&$ 0.84 \pm 0.01 $\\
500--1000   &$ 1.02 \pm 0.01  $&$1.09\pm0.004$ &$ 1.28  \pm 0.04 $&$ 0.87 \pm 0.01 $\\
1000--2000  &$ 1.15 \pm 0.02  $&$1.22\pm0.007$ &$ 1.50  \pm 0.07 $&$ 0.98 \pm 0.02 $\\
2000--4000  &$ 1.35 \pm 0.04  $&$1.39\pm0.014$ &$ 1.54  \pm 0.11 $&$ 1.11 \pm 0.04 $\\
4000--10000 &$ 0.96 \pm 0.07  $&$0.92\pm0.022$ &$ 0.98  \pm 0.15 $&$ 0.89 \pm 0.07 $\\
10000--20000&$ 0.98 \pm 0.04  $&$0.81\pm0.058$ &$ 0.62  \pm 0.39 $&$ 0.91 \pm 0.21 $\\
20000--50000&   \nodata        &$0.87\pm0.16 $ &$ 1.14  \pm 1.3  $&$ 0.71 \pm 0.73 $\\
\enddata
\tablecomments{``Total'' refers to fitting sum of all model diffuse components, 
``IC'' and ``gas'' refer to fits separating IC from gas-related components. 
Only statistical errors are given.\vspace{-3mm}}
\end{deluxetable*}

\placetable{estimates_of_EB}
 

\begin{deluxetable*}{cccccc}[!t]
\tablecolumns{6}
\tablewidth{0mm}
\tabletypesize{\footnotesize}
\tablecaption{Estimates of \EB\ obtained by fitting optimized model 500190 to 
EGRET data.
\label{estimates_of_EB}}
\tablehead{
\colhead{} &
\colhead{$10^\circ<|b|<80^\circ$,} &
\colhead{Adopted {\it fractional}} &
\multicolumn{2}{c}{All-sky} \\ \cline{4-5}
\colhead{Energy, MeV} &
\colhead{total} &
\colhead{systematic error} &
\colhead{Total} &
\colhead{IC+gas} &
\colhead{\citet{sreekumar98}\tablenotemark{a}} 
}
\startdata
30--50      &$ 16.8 \pm 0.66$ &0.30&$ 16.0 \pm  0.48$  &$17.0\pm0.84$     &$24.0 \pm 7.0 $  \\
50--70      &$ 10.6 \pm 0.19$ &0.15&$ 10.4 \pm  0.13$  &$11.1\pm0.22$     &$13.26 \pm 2.58 $  \\
70--100     &$ 6.66 \pm 0.10$ &0.10&$  6.7 \pm  0.065$ &$ 6.75\pm0.11$    &$ 7.83 \pm 1.05$  \\
100--150    &$ 4.48 \pm 0.07$ &0.10&$  4.6 \pm  0.045$ &$ 4.55\pm0.076$   &$ 5.5  \pm 0.75$  \\
150--300    &$ 3.92 \pm 0.06$ &0.10&$  4.2 \pm  0.040$ &$ 3.92\pm0.067$   &$ 5.4  \pm 0.72$  \\
300--500    &$ 1.20 \pm 0.04$ &0.10&$  1.39\pm  0.025$ &$ 1.17\pm0.041$   &$ 1.97 \pm 0.268$  \\
500--1000   &$ 0.76 \pm 0.04$ &0.10&$  0.93\pm  0.023$ &$ 0.70\pm0.037$   &$ 1.36 \pm 0.185$  \\
1000--2000  &$ 0.32 \pm 0.03$ &0.10&$  0.39\pm  0.018$ &$ 0.26\pm0.028$   &$ 0.617\pm 0.084$ \\
2000--4000  &$ 0.24 \pm 0.02$ &0.10&$  0.30\pm  0.014$ &$ 0.25\pm0.022$   &$ 0.304\pm 0.044$ \\
4000--10000 &$ 0.117\pm 0.02$ &0.10&$  0.13\pm  0.010$ &$ 0.12\pm0.016$   &$ 0.1956\pm 0.0288$ \\ 
10000--20000&$ 0.025\pm 0.04$ &0.25&$  0.034\pm 0.011$ &$ 0.04 \pm0.02$   &$ 0.053\pm 0.016$ \\
20000--50000&   \nodata       &0.25&$  0.011\pm 0.022$ &   \nodata        &$ 0.0276\pm 0.0096$ \\
50000--120000&  \nodata       &\nodata& \nodata        &   \nodata        &$ 0.0147\pm 0.0063$ 
\enddata
\tablecomments{``Total'' refers to fitting sum of all model diffuse components, 
``IC'' and ``gas'' refer to fits separating IC from gas-related components.
For our fits statistical errors are given together with adopted systematic errors
on the EGRET response.
Units: 10$^{-6}$ cm$^{-2}$ sr$^{-1}$ s$^{-1}$.}
\tablenotetext{a}{Values are from their Table~1 integrated over the energy bin.
The error bars include systematic errors. \vspace{-5mm}}
\end{deluxetable*}

\section{Determining the \EB \label{sectionEB}} 
The scaling factors determined for the fits in \regionG\
(Table~\ref{scaling_factors_EB}) reflect the deviations from the
model, visible in Fig.\ 8 of \citet{SMR04}, and are typically between
0.85 and 1.15 which is satisfactory  considering the EGRET systematic
uncertainty $\sim$15\% indicating a
good agreement outside the plane.  There is only one region
(2--4 GeV) where the scaling factor is as large as 1.35.  
For the all-sky fits the 1--2 and
2--4 GeV factors are large in the case of the IC component,
probably related to an underestimate of the ISRF or of
the electron density.  In the range 30--50 MeV, the gas-related factor
is 1.31 which may indicate underestimated bremsstrahlung,
but the EGRET data in this region are subject to a correction as a
result  of instrumental calibrations (\citealt{Thompson1993b}; see
also discussion of the Kniffen factor in Section 3.1 of \citealt{SMR04})
and thus more uncertain compared to other energies.  These are the
least reliable ranges of our \EB\ determination reflecting the
discrepancy in the spectrum mentioned in Section 6 of \citet{SMR04}.
The \EB\ is however not very sensitive to the scaling factor, since
the intercept requires only a small extrapolation to zero Galactic
flux.  The observed versus (fitted) predicted intensities are shown in
Fig.~\ref{gamma_pred_obs_500190} (region  $|b|>10^\circ$) using the
parameters  in  Table~\ref{scaling_factors_EB} and
Table~\ref{estimates_of_EB}.  These plots give an idea of the
reliability of the analysis: the relation is linear to good accuracy
and the \EB\ well determined.

Table~\ref{estimates_of_EB} presents the \EB,
integrated over each energy bin since this is given
directly by the fitting. The values from \citet{sreekumar98} (their
Table 1) were thus integrated (assuming an $E^{-2}$ spectrum) to
compare with ours. It is seen that the two fitted regions 
($|b|>10^\circ$ and all-sky) give
consistent results, indicating that there is no large systematic
effect; it shows a model-dependent systematic uncertainty of
5--25\%. This is comparable to the $\sim$15\% systematic uncertainty
on EGRET effective area. The adopted energy-dependent systematic 
error for EGRET data is also given in the Table. The total error
is obtained by combining the statistical and EGRET systematic 
errors in quadrature. 

\begin{figure}[!t]
\centering 
\includegraphics[width=\fwa]{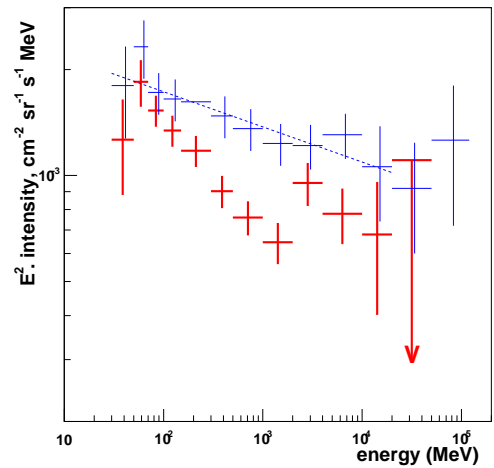}

\caption{Comparison of our \EB\ spectrum (solid, red)
as given in Table \ref{EGRB_flux} with that
from \citet{sreekumar98} (dots, magenta). 
The dashed (blue) line is the \citeauthor{sreekumar98} fit:
$2.743\times10^{-3} E^{-2.1}$ cm$^{-2}$ s$^{-1}$ sr$^{-1}$ MeV$^{-1}$. 
\vspace{0mm}}
\label{spectrum_isotropic}
\end{figure}

Fig.~\ref{spectrum_isotropic} shows the spectrum of \EB\ as derived in
the present work (Table~\ref{EGRB_flux}).  
The data points of the \emph{differential} spectrum
were obtained from Table~\ref{estimates_of_EB} (columns 1--3 
and column 4 above 10 GeV) assuming a
power-law spectrum within the energy bin. The two $E>10$ GeV points
were taken as given in ``All-sky, Total'' column, while the 20--50 GeV
point was plotted as a $1\sigma$ upper limit. Above 50 GeV the data
were insufficient to give a significant result by our method. 

\placefigure{spectrum_isotropic}

\begin{figure*}
\centering 
\includegraphics[height=\fwb,angle=90]{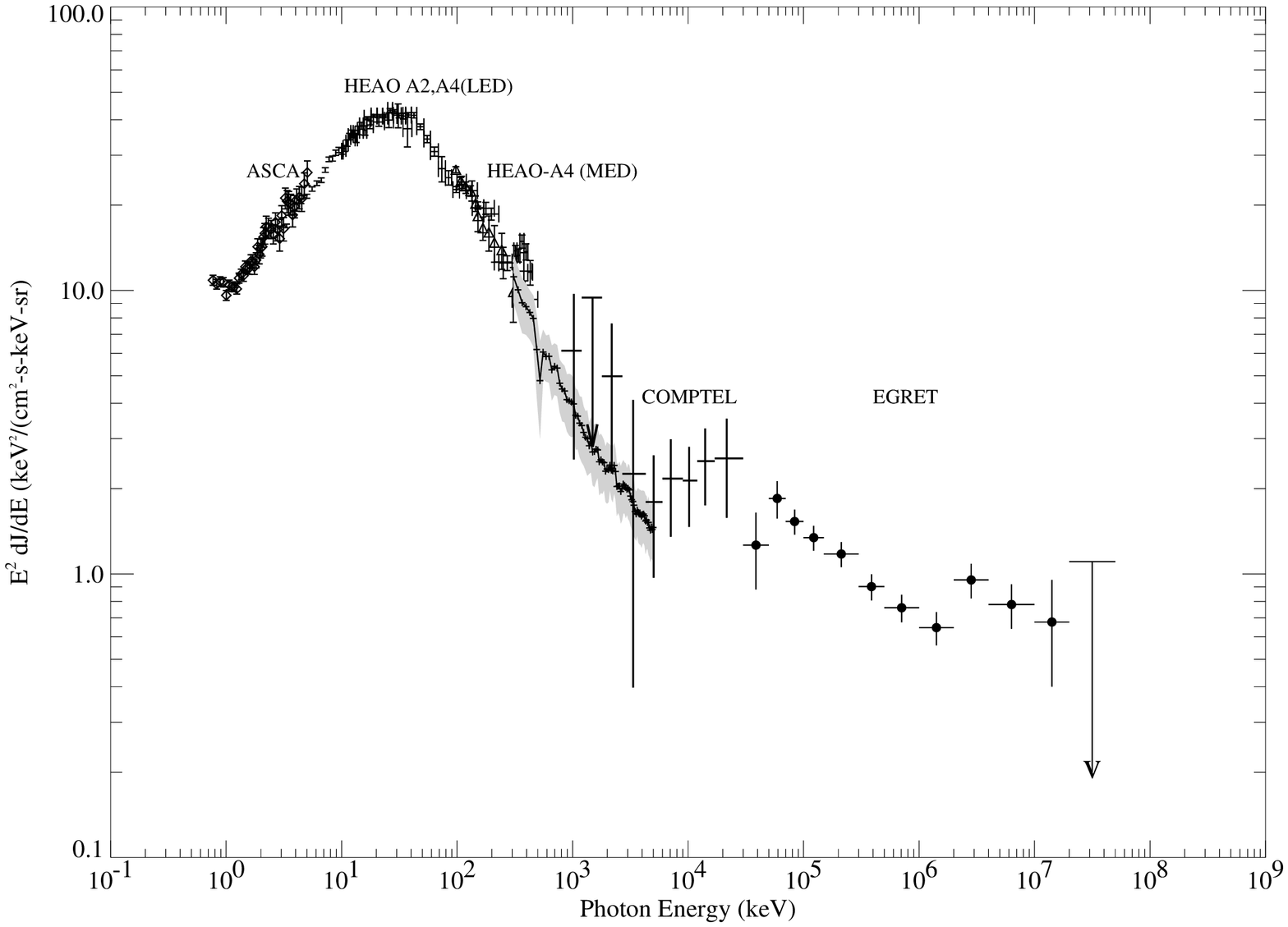}

\caption{Extragalactic X-ray and \gray spectrum. Data compilation from
\citet{sreekumar98}  except for COMPTEL \citep{weidenspointner00} and
EGRET 30 MeV -- 20 GeV (this work). }
\label{spectrum_EGRB}
\end{figure*}

\placefigure{spectrum_EGRB}

\placetable{EGRB_flux}

\begin{deluxetable}{ccc}
\tablecolumns{3}
\tablewidth{0mm}
\tabletypesize{\footnotesize}
\tablecaption{\EB\ intensity.
\label{EGRB_flux}}
\tablehead{
\colhead{} &
\colhead{Intensity,} &
\colhead{}\\
\colhead{$E_\gamma$, MeV} &
\colhead{cm$^{-2}$ sr$^{-1}$ s$^{-1}$ MeV$^{-1}$} &
\colhead{Total error}
}
\startdata
30--50      &$ 8.40\times10^{-7~}$&$ 2.54\times10^{-7~}$\\
50--70      &$ 5.30\times10^{-7~}$&$ 0.80\times10^{-7~}$\\
70--100     &$ 2.22\times10^{-7~}$&$ 0.22\times10^{-7~}$\\
100--150    &$ 8.96\times10^{-8~}$&$ 0.91\times10^{-8~}$\\
150--300    &$ 2.61\times10^{-8~}$&$ 0.26\times10^{-8~}$\\
300--500    &$ 6.00\times10^{-9~}$&$ 0.63\times10^{-9~}$\\
500--1000   &$ 1.52\times10^{-9~}$&$ 0.17\times10^{-9~}$\\
1000--2000  &$ 3.20\times10^{-10}$&$ 0.44\times10^{-10}$\\
2000--4000  &$ 1.20\times10^{-10}$&$ 0.16\times10^{-10}$\\
4000--10000 &$ 1.95\times10^{-11}$&$ 0.30\times10^{-11}$\\ 
10000--20000&$ 3.40\times10^{-12}$&$ 1.39\times10^{-12}$\\
20000--50000&$<1.11\times10^{-12}$& \nodata\\
\enddata
\tablecomments{\EB\ intensity (as plotted in 
Fig.~\ref{spectrum_isotropic}) derived from columns 1--3 of 
Table~\ref{estimates_of_EB} and column 4 above 10 GeV.}
\end{deluxetable}

Our estimates lie significantly below those of \citet{sreekumar98}, in
most energy ranges.  The positive curvature in our \EB\ spectrum is
interesting and not unexpected for a blazar origin \citep{salamon98}
but in view of the systematic uncertainties should not be taken too
literally; a similar, less pronounced effect is present in the
\citeauthor{sreekumar98} spectrum.   Our spectrum is not consistent
with a power law.  The 0.1 -- 10 GeV intensity for the region outside
the Galactic plane (\regionG) of $(11.1\pm0.1)$\intensityunits can be
compared with $(14.5\pm0.5)$\intensityunits from \citet{sreekumar98}.

Fig.~\ref{spectrum_EGRB} shows the extragalactic X- and \gray
background, using the compilation by \citet{sreekumar98} but  using our
new EGRET values, and also updated COMPTEL results
\citep{weidenspointner00}.

\section{Further checks for systematics due to model}
In order to check the robustness of our \EB\ estimates, we compare
estimates based on data in different hemispheres and four quarter
spheres (Table~\ref{isotropy}).  This tests for the presence of
apparent anisotropies in the \EB\ values, and thus gives an
independent estimate of the systematic error due to model
inadequacies.  In order to have sufficient statistics we consider only
the integral energy range 0.1--10 GeV.  The hemisphere deviations are up
to 16\% from the fit to the full $|b|>10^\circ$ region.  The quarter
sphere values deviate by up  to 25\% from the full region. This may
reflect essentially different exposures (e.g., southern hemisphere is
less exposed than northern hemisphere).  The deviations are also
probably indicating lack of symmetry in the observed Galactic \gray
emission not reflected in the model. However, using all regions
together,  as we have done for our results, will tend to compensate
such systematics.

\placetable{isotropy}


\begin{deluxetable*}{clcl}
\tablecolumns{4}
\tablewidth{0mm}
\tabletypesize{\footnotesize}
\tablecaption{Test for model-dependent systematics.
\label{isotropy}}
\tablehead{
\colhead{$l$} &
\colhead{$b$} &
\colhead{Intensity 0.1--10 GeV} &
\colhead{Description}
}
\startdata
0--360       &$<-10,>+10$  &$ 11.10\pm 0.12$ & N+S hemispheres          \\
0--360       &$<-10     $  &$ 11.70\pm 0.15$ & N hemisphere             \\
0--360       &$>+10     $  &$ ~9.28\pm 0.21$ & S hemisphere             \\  
270--90      &$<-10,>+10$  &$ 11.90\pm 0.17$ & Inner Galaxy N+S         \\
90--270      &$<-10,>+10$  &$ ~9.75\pm 0.17$ & Outer Galaxy N+S         \\
0--180       &$<-10,>+10$  &$ 10.80\pm 0.17$ & Positive longitudes N+S  \\
180--360     &$<-10,>+10$  &$ 11.60\pm 0.16$ & Negative longitude N+S   \\
270--90      &$>+10     $  &$ 13.00\pm 0.22$ & Inner Galaxy N           \\
270--90      &$<-10     $  &$ ~9.14\pm 0.32$ & Inner Galaxy S           \\
90--270      &$>+10     $  &$ 10.60\pm 0.22$ & Outer Galaxy N           \\
90--270      &$<-10     $  &$ ~8.18\pm 0.34$ & Outer Galaxy S           \\
\enddata
\tablecomments{Units: $10^{-6}$ cm$^{-2}$ sr$^{-1}$ s$^{-1}$. 
Errors are statistical only.}
\end{deluxetable*}

\begin{figure}[t]
\centering 
\includegraphics[width=\fwa]{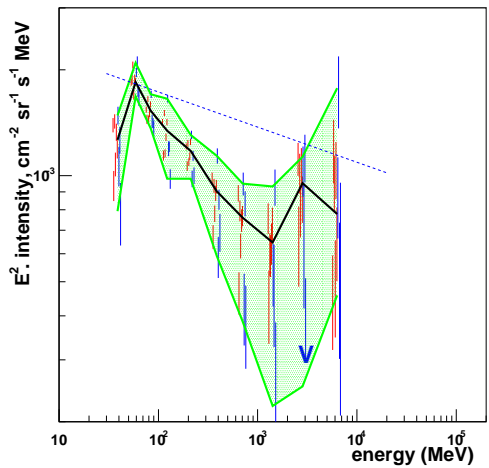}

\caption{\EB\ spectrum determined for each of the regions in 
Table~\ref{isotropy}.
Red bars: hemispheres, blue bars: quarter spheres. 
The upper and lower bound spectra are shown in green, and the adopted \EB\ in black. 
Energies are shifted slightly for clarity.
The dashed (blue) line is the \citeauthor{sreekumar98} fit.
}
\label{spectrum_limits}
\end{figure}

Fig.~\ref{spectrum_limits} shows the \EB\ determined from the fits to
all the regions of Table~\ref{isotropy}. The quarter-sphere spectra
show more scatter than the hemisphere spectra, as expected since there
are less data fitted. The extrema are plotted to give an indication of
the extreme upper and lower bounds on the spectrum accounting for
systematics due to the model.  Compared with the  adopted \EB\ (also
shown), the extreme upper limits lie 20-50\% higher (100\% for 4--10
GeV), the lower limits 10-50\% lower.  These extreme bounds are quite
conservative; note that since this plot illustrates the
model-dependent error,  the additional error due to exposure
uncertainty (Section 3) is not included here.  The adopted spectrum
using all the $|b|>10^\circ$ data is expected to be the most robust
against such systematics since large-scale asymmetry effects will tend
to cancel.

\section{Conclusions}
Based on our new optimized model for the diffuse Galactic \gray
emission, a new \EB\ spectrum has been derived. It is lower and
steeper than found by \citet{sreekumar98}; it is not consistent with a
power-law, and shows some positive curvature as expected for an origin
in blazars.

\begin{acknowledgements}
We would like to particularly thank David Bertsch for assistance and
discussions on the subject of the events and instrumental response of
the EGRET telescope above 10 GeV and Seth Digel for providing the
kinematically  analysed \hi\ and CO data used in this work.  A part of
this work has been done during a visit of Igor Moskalenko to the
Max-Planck-Institut f\"ur extraterrestrische Physik in Garching; the
warm hospitality and financial support of the Gamma Ray Group is
gratefully acknowledged. Igor Moskalenko  acknowledges partial support
from a NASA Astrophysics Theory Program grant.  Olaf Reimer
acknowledges support from the BMBF through DLR grant QV0002.
\end{acknowledgements}






\end{document}